\documentclass[12pt]{iopart}
\usepackage{epsf}
\begin{document}
\title{Quasi-stationary binary inspiral II:
  Radiation-balanced boundary conditions}
\author{John T Whelan\dag\footnote[2]{Electronic Address:
    whelan@itp.unibe.ch}, William Krivan\S\footnote[4]{Current
    Address: Center for Genomics Research, Karolinska Institutet,
    Berzelius v\"{a}g 37, 17177 Stockholm, Sweden}
  and Richard H Price\S}
\address{\dag Institut f\"{u}r theoretische Physik, Universit\"{a}t Bern,
  Sidlerstrasse 5, CH-3012 Bern, Switzerland}
\address{\S Department of Physics, University of Utah,
         Salt Lake City, UT 84112}

\begin{abstract}
The quasi-stationary method for black hole binary inspiral is an
approximation for studying strong field effects while suppressing
radiation reaction. In this paper we use a nonlinear scalar field 
toy model (i) to explain the underlying
method of approximating binary motion by periodic orbits with
radiation; (ii) to show how the fields in such a model are
found by the solution of a boundary value problem; (iii) 
to demonstrate how a good approximation to the outgoing radiation
can be found by finding fields with a balance of ingoing and 
outgoing radiation (a generalization of standing waves).
\end{abstract}

\vspace{28pt plus 10pt minus 18pt}
     \noindent{\small\rm Submitted to: {\it  Class.\ Quantum Grav.}\par}
\maketitle

\section{Introduction and overview}\label{sec:intro}
The inspiraling of black hole binaries is receiving much recent
attention both because it is an exciting potential source of
detectable gravitational waves\cite{fh}, and due to its inherent
interest as a strong field gravitational interaction. The process of
inspiral and merger is being investigated with a number of techniques.
Newtonian and post-Newtonian computations\cite{PN} are appropriate to
the early stages of inspiral; numerical relativity\cite{numrel} and
black hole perturbation theory\cite{CL} are used for the late strong
field stage of the inspiral. The present paper deals with aspects of
the intermediate phase, when strong field effects are too important
for post-Newtonian methods to be useful, but in which the full power
of numerical relativity is not required.  Detweiler and
collaborators\cite{det} have drawn attention to the possible value of
an approximation based on the comparison of the orbital time
$\tau_{orb}$ for the binary holes with the time $\tau_{rad}$ on which
gravitational radiation acts to change the orbit.  {From} a
dimensional analysis of an equal mass binary of mass $M$ and
separation $a$
this ratio is
\begin{equation}\label{tauratio}
\frac{\tau_{orb}}{\tau_{rad}}
\propto\left(
\frac{
GM/c^{2}}
{a}
\right)^{5/2}\ .
\end{equation}
The factor $GM/(ac^{2})$ on the right is an indicator of ``how
relativistic'' the gravitational interaction is between the two holes.
When $a$ is on the order of 30 or so times $GM/c^{2}$, the ratio in
(\ref{tauratio}) will be small, and the orbits may be quasiperiodic,
but nonradiative relativistic effects may still be important.

It is useful to focus on a particular strong field relativistic effect
of great importance: the innermost stable circular orbit (ISCO). The
particle limit in gravitation theory treats the mass $\mu$ of one of
the orbiting objects as much smaller than $M$, the mass of the
other. In this approximation, there is a minimum radius for circular
orbital motion, at a value $r=6GM/c^{2}$ in the case of a nonrotating
hole.  The existence of this limiting orbit is a purely relativistic
effect; there is no such limit in Newtonian theory. It is,
furthermore, unrelated to radiation. In the particle limit radiation
reaction is a force of order $(\mu/M)^{2}$ and is ignored; the
particle moves on a geodesic.  There is no such justification for
neglecting radiation reaction for the case of binary motion of equal
mass holes. Since all orbits are being degraded by gravitational
radiation, there is no meaning to ``stability'' in principle. But
there is an important practical question: Do relativistic forces arise
in the late binary motion that drive orbiting particles to plunge
inward on a time scale much shorter than the time scale due to
radiation?  It is very plausible that this sort of ``practical''
meaning can be given to the question of stability. For radial infall
of equal mass holes, we know\cite{misprob} that the radiated energy is
a very small fraction of the mass energy of the system as the holes
fall into each other from moderate separations. Gravitational
radiation reaction, therefore, cannot be a significant modification of
the motion of the holes.

This and other questions can be investigated in the absence of
radiation reaction by seeking an approximation to the slowly evolving
spacetime which is periodic, or, in the case of circular orbits,
stationary.  It turns out that this greatly changes the mathematical
nature of the solution process. The problem of evolving Cauchy data is
converted into a boundary value problem. Past experience with this
type of problem indicates that this boundary value problem suffers
from none of the instability difficulties of numerical nonlinear
evolution.  The investigation of such questions has usually been based
on an {\it ansatz}\cite{Cooknorad,WMM} for suppressing radiative
degrees of freedom in general relativity (GR), but such approaches by
necessity only solve a subset of the full Einstein equations.  We
propose a very different approach.  We do not attempt to suppress
radiation fields, rather we suppress radiation reaction by requiring
that the energy lost to outgoing radiation be replaced by a
corresponding amount of ingoing radiation.

Since the goals of this paper are to introduce the general ideas
behind a new approximation scheme, and to demonstrate the numerical
implementation of this scheme, we devote our attention here to a toy
model rather than to GR, with its added complexities.  The toy model
is the simplest system containing the relevant features of radiation,
nonlinearity, and decaying orbits, namely that of a nonlinear scalar
field in 2+1 dimensions.  The choice of two rather than three spatial
dimensions is made for two reasons: First and foremost, it makes the
problem of finding a stationary solution less computationally
intensive, since the wave equation has to be solved on a two- rather
than three-dimensional grid.  Second, a scalar field theory in 2+1
dimensions is equivalent to the same theory in 3+1 dimensions with all
the sources and fields required to be translationally invariant in one
spatial dimension.  This theory with line-like sources is analogous to
the problem of orbiting line-like sources in 3+1 GR. (Note that the
scalar theory in 2+1 dimensions is {\em not} analogous to a problem in
2+1 GR. In 2+1 GR, gravity, i.e., Riemann curvature, vanishes outside
gravitating bodies.)  The formalism for 3+1 line-like sources in GR
has already been developed\cite{paperI}, and the spacetime they
generate will be the topic of a subsequent paper.

The analogy, with either point-like or line-like sources, between the
scalar field problem and full GR is of course only a qualitative one,
and we consider the scalar field results only as a test of the general
method and not in any way a realistic model of the problem in GR.  By
the same token, the study now in progress of orbiting line sources in
GR will not serve as a realistic simulation of the astrophysical
problem with localized sources, but rather as a toy model that
incorporates the added complexity of a gravitational problem while
remaining computationally similar.  Line-like sources in GR have
features not found with localized sources, such as the lack of an ISCO
for test particles and the lack of an asymptotically flat region in
which gravitational waves have familiar properties\cite{cyl}. On the
other hand, the analogy between the 2+1 and 3+1 problems in scalar
field theory is much closer, as illustrated analytically in
\ref{app:threeD} in the absence of nonlinearities.  We therefore
consider the difference between looking numerically at 2+1 rather than
3+1 \emph{scalar} field theory to be one primarily of computational
complexity.  Finally,

Our toy model consists of two particles, each with a charge that
couples to the nonlinear scalar field. To understand our method it is
useful to consider three different solutions for the particle motion
and the scalar field: I\@. The radiation is outgoing, and as a result
of the loss of energy due to the radiation, the orbiting particles
spiral inward. II\@. The radiation is outgoing, but due to
constraining forces the particles remain in circular periodic
orbits. III\@.  Again, the particles move in circular periodic orbits,
but now the scalar radiation is balanced and the waves are some
generalization of standing waves; there is as much ingoing radiation
energy flux as outgoing.  The solution of type I is the (scalar field,
2+1) analog of the problem of binary inspiral in GR.  The solution of
type II is a reasonable approximation of the type-I problem when
$\tau_{orb}/\tau_{rad}\ll1$. This type of solution makes sense for a
scalar field model; the constraining forces that maintain the periodic
motion can be invoked {\em ad hoc}. Such forces need not couple to the
scalar field, so they can be specified to have  no effect on the
problem other than to maintain the periodic circular orbits. In GR, on
the other hand, all interactions couple to gravitation, and a solution
of type II does not make sense. Solutions with periodic orbits and
outgoing radiation should therefore not exist in GR. Solutions of type
III, however, are not {\em a priori} ruled out by such considerations.

There are two principal goals of this paper. The first is to show that
the type-II fields can be found by solving a boundary value problem,
and to suggest that such a problem is much more easily solved than a
Cauchy evolution problem.  Our second goal is to demonstrate that a
solution of the type-III problem can also be computed without
evolution, and that the computed solution gives a good approximation
to the fields of the type-II problem, when the conditions of our
approximation are valid. The implication is that the physical problem
in GR, the type-I problem, can be approximated from the ``easily''
solved type-III problem.
 
The remainder of this paper will be organized as follows. The basic
toy model, a nonlinear scalar field will be introduced in
Section~\ref{sec:toy} and solutions, both analytic and numerical, will
be presented for periodic fields with outgoing boundary conditions.
In Section~\ref{sec:balanced} we discuss the meaning of radiation
balanced fields and we make a particular choice (the ``TSGF'') of such
fields for our nonlinear model.  Issues related to the application of
the quasi-stationary approximation to GR are discussed in
Section~\ref{sec:disc}.  \ref{app:gfsoln} gives some technical details
of the general solution to the linear wave equation with equal flux of
ingoing and outgoing radiation. \ref{app:threeD} presents the solution
to a linear scalar field problem
in 3+1 dimensions, as an illustration of the relationship of the 3+1
and 2+1
problems.  Finally, \ref{app:force} demonstrates explicitly (in the
linear case) the vanishing of the radiation reaction force for the
radiation-balanced solutions.

\section{Nonlinear scalar field with periodic orbits}\label{sec:toy}
Our toy model is based on a nonlinear field $\psi$ in 2+1
dimensions satisfying
\begin{equation}\label{gennonlin}
\nabla^{2}\psi-\partial_{t}^{2}\psi+\lambda{\cal F}(\psi,\rho)
=-\sigma\ .
\end{equation}
where $\sigma$ is a scalar source for the field.  The spacetime for
the field is Minkowskian and we use polar
spatial coordinates
$\rho,\phi$.  The term $\lambda{\cal F}(\psi,\rho)$ is a term nonlinear in
$\psi$ that  may be an explicit
function of $\rho$, but not an explicit function of $\phi$ or $t$.  
For later computational convenience 
we require that the nonlinear term 
satisfy the symmetry
condition
\begin{equation}\label{piflip}
{\cal F}(\psi,\rho)=-{\cal F}(-\psi,\rho)\ .
\end{equation}
The constant  $\lambda$ is a parameter that governs the strength of 
the nonlinear term.
We now specialize to the 
source
\begin{equation}\label{lines}
\sigma=a^{-1}Q\delta(\rho-a)
\left[
  \delta(\phi-\Omega t-\pi/2)-\delta(\phi-\Omega t-3\pi/2)
\right]
\end{equation}
representing two
point sources (in two spatial dimensions) of scalar charge density
$\pm Q$, moving around each other in circular orbits of radius $a$, at
angular speed $\Omega$.

If we were to treat (\ref{gennonlin}) as an evolution problem, we
would have to specify Cauchy data and integrate forward in time to
find $\psi(\rho,\phi,t)$. Instead we concern ourselves only with the
steady state solution, a solution that would evolve after transients
associated with the initial conditions are radiated away. This steady
state solution would embody the symmetries of the source.  In
particular, the source depends on $\phi$ and $t$ only in the
combination ${\varphi}\equiv \phi-\Omega t$ and we seek a solution
with the same symmetry. That is, we look for a solution
$\psi(\rho,{\varphi})$. Since the left hand side of (\ref{gennonlin})
does not have an explicit dependence on $\phi$ or $t$, such a solution
is allowed.

When (\ref{gennonlin}) is restricted to solutions of the form
$\psi(\rho,{\varphi})$ it reduces to
\begin{eqnarray}
\fl\frac{1}{\rho}\frac{\partial}{\partial \rho}\left(
\rho\frac{\partial\psi}{\partial\rho} \right)+\left[
\frac{1}{\rho^{2}}-\Omega^{2}
\right]\frac{\partial^{2}\psi}{\partial{\varphi}^{2}}+\lambda{\cal
F}(\psi,\rho)=-\sigma(\rho,{\varphi})\nonumber\\
=a^{-1}Q\delta(\rho-a)
\left[
  \delta({\varphi}-3\pi/2)-\delta({\varphi}-\pi/2)
\right]
\ .
\label{reducedwveq}
\end{eqnarray}

\begin{figure}[ht]
  \begin{center}
\epsfxsize=.5\textwidth \epsfbox{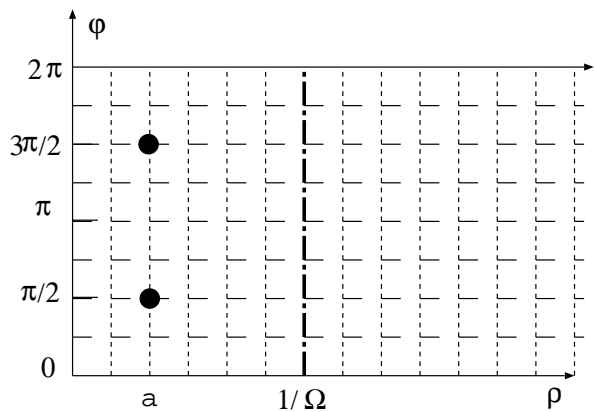}
  \end{center}
\caption{Coordinate regions for the periodic wave equation.
  The sources are shown as dark circles. The light cylinder is shown
  as the line with dots and dashes at $\rho=1/\Omega$.  The outer
  boundary, $\rho=\rho_{\rm max}$, is not shown.}
\label{coordgrid}
\end{figure}
The coordinate regions for this equation are shown in
Figure~\ref{coordgrid}. Although this figure misrepresents the topology
(a disk) of the physical problem, it is the appropriate description of
the ($\rho,{\varphi}$) grid on which (\ref{reducedwveq}) is to be solved
as a finite difference equation. The specific conditions to be used in
solving the problem are the following: 
(i) Due to the symmetry of our source, and to condition
(\ref{piflip}), the solution is to have the symmetry
$\psi(\rho,{\varphi})= -\psi(\rho,{\varphi}+\pi)$. This allows us to
restrict the numerical solution to the range $0<{\varphi}<\pi$.
(ii) Due to the antisymmetry under
${\varphi}\rightarrow{\varphi}+\pi$, the solution for $\psi$ must
vanish at all $\rho=0$ grid points.
(iii) Outgoing Sommerfeld boundary conditions
are imposed at some maximum radius ${\rho_{\rm max}}$ by requiring
\begin{equation}\label{outgoing}
\left.\left(\frac{\partial\psi}{\partial\rho}
-\Omega\frac{\partial\psi}{\partial{\varphi}}\right)
\right|_{\rho={\rho_{\rm max}}}=0
\ .
\end{equation}
If (\ref{reducedwveq}) and the above conditions (i-iv) are put on 
a grid of $N_{r}$ radial lines and $N_{{\varphi}}$ lines, containing 
$N\equiv N_{r}\times N_{{\varphi}}$ grid points with {\em a priori}
unknown values of $\psi$, a system of $N$ equations for these unknowns
is found. The solution of this system is the finite difference solution
of our physical problem.

The finite difference procedure just outlined is relatively
straightforward, but it has an unusual feature.  Note that the nature
of the differential equation in (\ref{reducedwveq}) changes at the
``light cylinder'' $\rho=1/\Omega$, shown as a line with dots and
dashes in Figure~\ref{coordgrid}. For $\rho<1/\Omega$ the equation is
formally elliptic, while for $\rho>1/\Omega$ it is hyperbolic.
Typically elliptic partial differential equations are solved as
boundary value problems, with auxiliary data given on a closed
boundary surrounding the region of the solution, but hyperbolic
equations are given Cauchy data on an ``initial'' hypersurface.  The
common wisdom is that a hyperbolic problem with data specified on a
closed surface can have more than a single solution\cite{mathwalk}.
Despite this we treat the entire coordinate region
($0\leq{\varphi}<\pi, 0\leq r \leq {\rho_{\rm max}}$) as a boundary
value problem.  We have not attempted to give a rigorous proof that
the boundary value approach to (\ref{reducedwveq}) and
(\ref{outgoing}) is well posed, but several nonrigorous justifications
are worth mentioning: (i) Although nonuniqueness is a possible feature
for a hyperbolic equations with boundary values, whether or not a
particular problem suffers from this difficulty depends on details of
the problem, not only on whether it is hyperbolic. (ii) The physical
problem described by the boundary value problem appears to be well
posed.  (iii) Numerical solutions of the boundary value problem are
stable and insensitive to numerical grid size.

Here, as in the next section, it will be useful to separate the
complexities of nonlinearity from other issues. To do this we
temporarily set $\lambda$ to zero so that (\ref{reducedwveq}) becomes
a linear equation. The outgoing radiation solution for this problem is
easily found, with standard techniques, in the form of a series of
Bessel $J_{m}$ and Neumann $N_m$ functions. For $\rho\geq a$, this
series is
\begin{equation}\label{Bessum}
\fl\psi_{\rm out}=Q\sum_{m=1,3,5,\ldots{}}(-1)^{(m+1)/2}J_{m}(m\Omega a)
\left[
N_m(m\Omega\rho)\sin{m{\varphi}}-J_{m}(m\Omega\rho)\cos{m{\varphi}}
\right]\ .
\end{equation}
This solution shows, by example, that there are no hidden difficulties
in finding a solution to (\ref{reducedwveq}) in the case $\lambda=0$,
and hence there is no fundamental problem in using a boundary value
approach to solve a problem with outgoing radiation. It also
demonstrates explicitly that, at least in the $\lambda=0$ case, the
light cylinder $\rho=1/\Omega$ is not a special surface in the
problem.  Note that (\ref{Bessum}) gives the solution for waves that
are ``outgoing at infinity.''  Solving the linear problem for waves
that are ``outgoing'' at a finite radius, i.e.\,, for the boundary
conditions in (\ref{outgoing}), is almost as simple as for true
outgoing waves. For $\rho\geq a$, the solution is
\begin{equation}
  \label{outwall} \psi=\psi_{\rm
out}-Q\sum_{m=1,3,5,\ldots{}}(-1)^{(m+1)/2}J_{m}(m\Omega a)
J_{m}(m\Omega\rho)\,{\rm Re}\left(\gamma_{m}\rme^{i m{\varphi}}\right)\ ,
\end{equation}
where
\begin{equation}
\label{def:gamout}
\gamma_{m}=\left.-\frac{H^{(1)}_{m}(z)+i \frac{d}{dz}H^{(1)}_{m}(z)}
{J_{m}(z)+i \frac{d}{dz}J_{m}(z)}\right|_{z=m\Omega {\rho_{\rm max}}
}\ .
\end{equation}
Here $H^{(1)}_{m}$ indicates the Hankel function of type 1. From a
well known property of Hankel functions\cite{abramstegun} the
numerator of (\ref{def:gamout}) vanishes as $m\Omega {\rho_{\rm
    max}}\rightarrow\infty$. This demonstrates that (aside from
roundoff and truncation error) a solution on a grid of finite radial
extent approaches the true outgoing solution as the radial extent of
the grid becomes infinite.  We have also checked that there are no
difficulties, with truncation or otherwise, in solving
(\ref{reducedwveq}) with finite difference methods. The solution found
in this way was compared with (\ref{Bessum}).  The agreement was
excellent, and the expected second order convergence of the finite
difference scheme was confirmed.

To investigate a nonlinear model numerically a specific choice must be
made for the nonlinear term ${\cal F}$. To suit our purposes this term
must satisfy several criteria in addition to the symmetry condition in
(\ref{piflip}). One criterion is that this nonlinear term be small
enough at the outer boundary ${\rho_{\rm max}}$, so that the outgoing
condition (\ref{outgoing}) is a good approximation at a large finite
radius.  The nature of this requirement can be seen if we choose
${\cal F}$ simply to be $\psi$, so that the nonlinear term in
(\ref{reducedwveq}) becomes $\lambda\psi$, and (\ref{reducedwveq}) is
the Helmholtz equation. The ``outgoing'' solution to this linear
problem is that given in (\ref{Bessum}), except that the arguments of
the Bessel and Neumann functions have $\sqrt{m^{2}\Omega^{2}+\lambda}$
in place of $m\Omega$. At large $\rho$ this solution will satisfy the
condition (\ref{outgoing}), for a particular angular Fourier mode, if
the $\Omega$ is replaced by $\sqrt{m^{2}\Omega^{2}+\lambda}\,/m$.  The
standard outgoing condition is then a good approximation only if
$|\lambda|\ll \Omega^{2}$.  To get a rough idea of the effect of
nonlinearities on boundary conditions we can view the nonlinear term
$\lambda{\cal F}$ as $\lambda_{\rm eff}\psi$ with $\lambda_{\rm eff}$,
the effective $\lambda$, taken to be $\lambda\langle{\cal
  F}\rangle/\left<\psi\right>$. Here ``$\langle\rangle$'' indicates
some sort of average (perhaps an r.m.s average over all ${\varphi}$
and one wavelength). A rough criterion for a nonlinear term that is
compatible with the boundary condition (\ref{outgoing}) is that
$\lambda_{\rm eff}$, at ${\rho_{\rm max}}$, be much smaller than
$\Omega^{2}$.  For our toy model to have interesting nonlinear
effects, however, there is a somewhat contradictory requirement: the
nonlinear term must be significant, even dominant, at small radius.  A
nonlinear term like ${\cal F}=\psi^{3}$ can be strong at small radii
and weak at large radii if the field $\psi$ falls off quickly enough.
For our line sources, the radiation fields fall off only as
$r^{-1/2}$. This slow fall off causes computational difficulties in
the application of boundary conditions.  For this reason we include a
factor $\exp{(-\left(\alpha\rho/a \right)^2)}$ in the nonlinear term.
To be certain that the nonlinear source is sufficiently well behaved
near the point sources at the points $(a,\pi/2)$ and $(a,3\pi/2)$, we
choose a source term that does not diverge at those points. The
specific choice used in our numerical investigations has been
\begin{equation}\label{gaussource}
{\cal F}=\exp(-(\alpha\rho/a)^{2})
\frac{\psi^3}{\left(\psi^{2}+1\right)^{3/2}}\ .
\end{equation}

\begin{figure}[ht]
\epsfxsize=3.3in\epsfbox{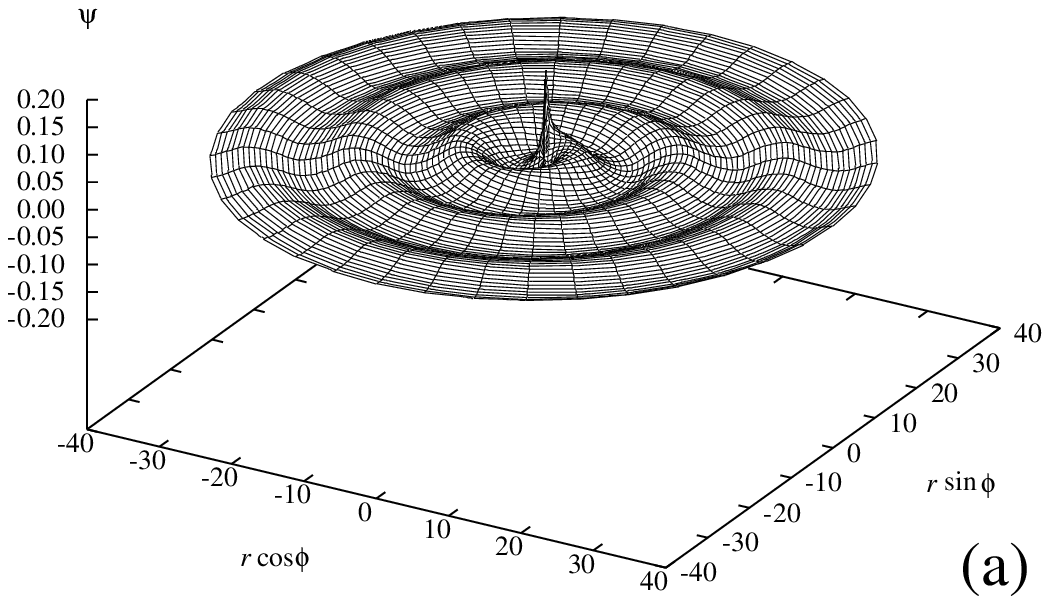}\hspace{-.6in}
\epsfxsize=3.3in\epsfbox{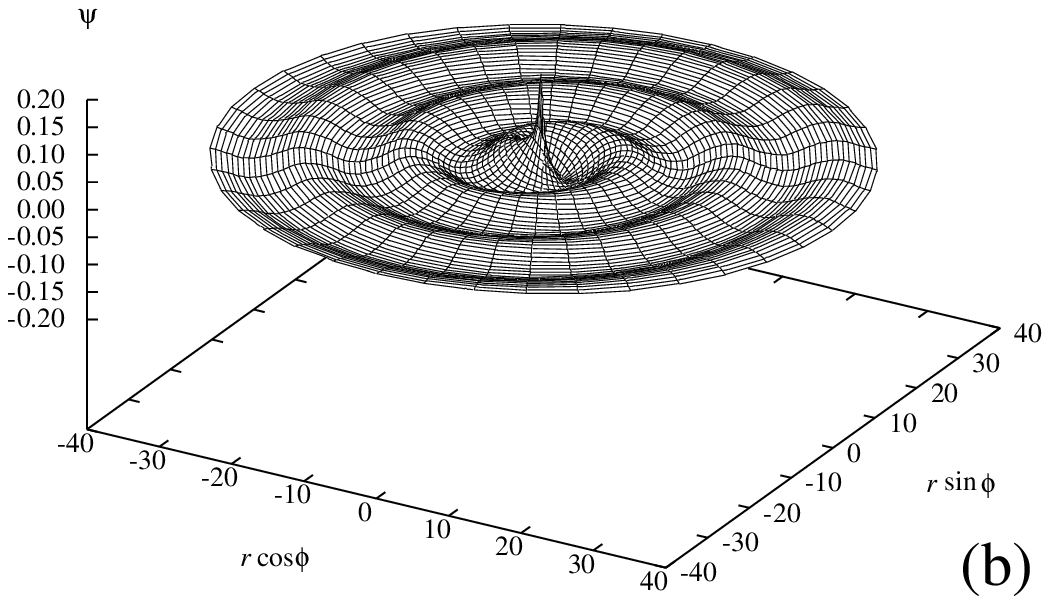}
\caption[psiout]{\label{psiout}
  The field $\psi$ as a function of $\rho\cos\phi$ and $\rho\sin\phi$
  for the two different times $\Omega t=0$ (a) and $\Omega t=\pi$ (b).
  This is a numerical solution to the nonlinear wave equation (with
  nonlinearity parameter $\lambda =1500$ and localization parameter
  $\alpha=1$) in the presence of outgoing-radiation boundary
  conditions.  The sources are located at a radius of $\rho=0.5$.
  Note that there are no discernable irregularities at the light
  cylinder $\rho=2$.}
\end{figure}

\begin{figure}[ht]
\hspace*{1.5in}\epsfxsize=3in \epsfbox{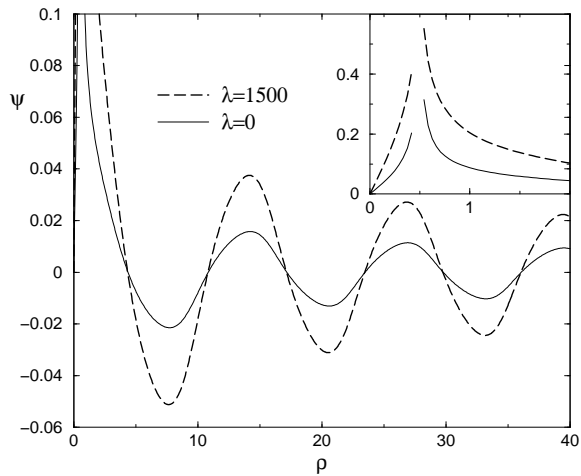}
\caption[psiout]{\label{inset} The linear (solid curve) outgoing
  solution for $\lambda=0$ and the nonlinear (dashed curve) outgoing
  solution for $\lambda=1500$, $\alpha=1$.  Both curves are for $Q=1$,
  $\Omega =0.5$, and $a = 0.5$. The fields are shown at
  ${\varphi}=\pi/2$.  The sharp peaks correspond to the location of
  the particle, at $\rho=0.5$.  Note that there are no discernable
  irregularities at the light cylinder $\rho=2$.}
\end{figure}

The outgoing wave pattern for $\psi$ is illustrated in
Figure~\ref{psiout}. The field $\psi$ is shown as a function of $\rho
\cos \phi$ and $\rho \sin \phi$ for two different times: (a) $\Omega
t=0$, and (b) $\Omega t=\pi$.  The figure illustrates how the rotation
of the ``rigid'' pattern sends waves outward.  The results were
obtained from numerical runs with $961 \times 41$ gridpoints in $\rho
\in [0,40]$ and ${\varphi} \in [0,\pi]$.  The parameter values are
$Q=1$, $\Omega =0.5$, $a = 0.5$, $\lambda =1500$, $\alpha=1$.  At
$\Omega t=0$, the positively charged particle is at $\phi=\pi/2$ and
at $\Omega t=\pi$ it is at $\phi=3 \pi/2$.
Figure \ref{inset} shows the importance of nonlinear effects for the
outgoing fields of Figure~\ref{psiout}. The field $\psi$ is shown as a
function of $\rho$ at ${\varphi}=\pi/2$, for $Q=1$, $\Omega =0.5$, and
$a = 0.5$.  The solid curve shows the solution of the linear problem
for $\lambda$ set to zero; the dashed curve shows the nonlinear fields
for $\lambda=1500$, $\alpha=1$.  Details of the fields near the source
particles are shown in the insert.  Though the nonlinear terms are
negligible in the wave zone, the amplitude of the waves is more than
doubled by the increase in the effective source strength produced by
the nonlinearity in the central region.

\section{Radiation-balanced boundary conditions}\label{sec:balanced}

The universality of gravitational coupling in GR means that periodic
orbits and outgoing radiation are not compatible.  In this section we
investigate a choice of solution for our toy model that {\em does}
seem applicable to periodic orbits in GR.  We introduce boundary
conditions that do not carry net energy away from the inner region of
the orbiting objects, boundary conditions containing equal measures of
ingoing and outgoing radiation.  There is a temptation to relate this
idea to that of ``standing waves,'' and to impose Dirichlet boundary
conditions that $\psi$ vanishes at some ``wall'' located at
$\rho={\rho_{\rm wall}}$. As in the previous section, it is useful to
separate issues of nonlinearity from other issues, and to look at the
$\lambda=0$ case first. For this linear problem the solution, for
$\rho>a$, with $\psi$ vanishing at $\rho={\rho_{\rm wall}}$ is
\begin{equation}
\label{standing} 
\fl\psi=
Q\sum_{m=1,3,5,\ldots{}}(-1)^{(m+1)/2}J_{m}(m\Omega a)
\left[
N_m(m\Omega\rho)+\beta_{m}J_{m}(m\Omega\rho)
\right]\sin{m{\varphi}}\ , 
\end{equation}
where 
\begin{equation}
\label{betastand}
\beta_{m}=-\frac{N_m(m\Omega {\rho_{\rm wall}})}
{J_{m}(m\Omega {\rho_{\rm wall}})}
\ .
\end{equation}
The expressions in (\ref{standing}) and (\ref{betastand}) do not
describe a well behaved solution. In general, the denominator for
$\beta_{m}$ will come arbitrarily close to zero, as larger and larger
values of $m$ are included in the sum. We have confirmed that the
finite difference solution to (\ref{standing}) and (\ref{betastand})
is not stable for small changes in the location of ${\rho_{\rm
    wall}}$, or for a change in the number of grid points. [Note that
increasing the number of angular grid divisions is roughly equivalent
to increasing the maximum value of $m$ included in the sum in
(\ref{standing}).]

In considering fields $\psi$ that are acceptable mixtures of ingoing
and outgoing waves, it is again helpful to look at the
linear ($\lambda=0$) problem.  In this case, a more-or-less obvious
acceptable choice of $\psi$ can be constructed simply by averaging the
solutions to (\ref{reducedwveq}) with ingoing waves at infinity and
with outgoing waves at infinity. The result, for $\rho>a$, is
\begin{equation}
\label{halfoutin}
\psi=Q\sum_{m=1,3,5,\ldots{}}(-1)^{(m+1)/2}J_{m}(m\Omega a)
N_m(m\Omega\rho)\sin{m{\varphi}}\ .
\end{equation}
Although this is not the most general solution of the linear problem
with equal amounts of in and outgoing waves, it is the most natural 
solution, as discussed in \ref{app:gfsoln}. 

The field in (\ref{halfoutin}) is a solution to (\ref{reducedwveq})
in the $\lambda=0$ case, but not for any simply stated boundary
conditions.  In particular, it does not correspond to the vanishing of
$\psi$ at some specific finite radius. Nor is it the limit in which
the ``wall'' of (\ref{standing}), (\ref{betastand}) is at $\infty$; no
such limit exists. Rather, it is a superposition of two solutions each
of which has a simply stated boundary condition (ingoing or
outgoing), and superposition is not valid for solutions of nonlinear
equations. We therefore reformulate our toy model so that fields can
be found that are the nonlinear equivalent of (\ref{halfoutin}). In
place of a partial differential equation, we introduce an integral
equation. The first step in doing this is to rewrite
(\ref{reducedwveq}) as
\begin{equation}\label{rewrite} 
{\cal L}\psi
=\sigma_{\rm eff}(\rho,{\varphi},\psi)\ ,
\end{equation}
where 
\begin{equation}\label{Ldef}
  {\cal L}\equiv
  -\frac{1}{\rho}\frac{\partial}{\partial\rho}
  \rho\frac{\partial}{\partial\rho}
  -\left(
    \frac{1}{\rho^{2}}-\Omega^{2}
  \right)\frac{\partial^{2}}{\partial{\varphi}^{2}}
\end{equation}
and
\begin{equation}\label{rhoeff} 
\sigma_{\rm eff}(\rho,{\varphi},\psi)\equiv\sigma(\rho,{\varphi})
+\lambda{\cal F}(\psi,\rho)
\ .
\end{equation}
We take the solution of (\ref{rewrite}-\ref{rhoeff}) to be 
the time symmetric Green function (TSGF) solution given by
\begin{equation}\label{GF} 
\psi_{\rm TSGF}(\rho,{\varphi})
=\int\int G_{\rm TS}(\rho,{\varphi};\rho',{\varphi}')
\sigma_{\rm eff}(\rho',{\varphi}'\psi(\rho',{\varphi}'))
\rho'\rmd\rho'\rmd{\varphi}'
\ ,
\end{equation}
where the Green function $G_{\rm TS}(\rho,{\varphi};\rho',{\varphi}')$
is the time symmetric inverse to the linear operator ${\cal L}$
corresponding to equal mixtures of ingoing and outgoing waves. In
principle, this Green function can be written explicitly as
\begin{eqnarray}
  \fl G(\rho,{\varphi};\rho',{\varphi}')_{\rm TS}=
  \cases{
    -\frac{1}{2\pi}\ln({\rho}/{\rho'})
    -\frac{1}{2}\sum_{m=1}^\infty
    J_m(m\Omega \rho')N_m(m\Omega\rho)\cos m({\varphi}-{\varphi}')
    & $\rho>\rho'$\\
    -\frac{1}{2}\sum_{m=1}^\infty
    N_m(m\Omega \rho')J_m(m\Omega\rho)\cos m({\varphi}-{\varphi}')
    & $\rho<\rho'$
    }
\nonumber\\
\label{GFexplicit} 
\end{eqnarray}
In practice, this series form of the time symmetric Green function is
not directly applicable to our numerical method, and it is more useful
to write the same radiation balanced solution with the following
symbolic notation. We denote the solution to the nonlinear problem
with outgoing boundary conditions as
\begin{equation}\label{outonly} 
\psi_{\rm out}={\cal L}^{-1}_{\rm out}\sigma_{\rm eff}\ ,
\end{equation}
where ${\cal L}^{-1}_{\rm out}$ is the inverse to ${\cal L}$ for
outgoing boundary conditions, i.e.\,, the retarded-time Green function.
The ingoing solution $\psi_{\rm in}$ is written in a parallel manner
using the advanced-time Green function ${\cal L}^{-1}_{\rm in}$.  The
linear superposition of the ingoing and outgoing (LSIO) solutions is
not itself a solution since the problem is nonlinear.

\begin{figure}[ht]
\hspace{1in}\epsfxsize=4in\epsfbox{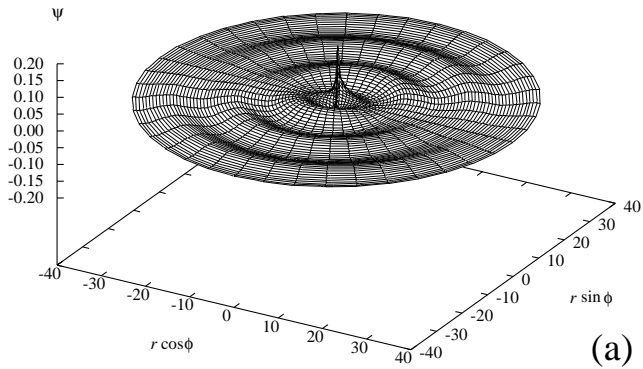}
\caption[psiout]{\label{standwvs}
  The field $\psi$ as a function of $\rho\cos\phi$ and $\rho\sin\phi$
  (at time $t=0$).  This is the numerical solution to the nonlinear
  wave equation (with nonlinearity parameter $\lambda=1500$ and
  localization parameter $\alpha=1$) according to the TSGF method of
  imposing a radiation-balanced boundary condition.  The sources are
  located at a radius of $\rho=0.5$.  Note that there are no
  discernible irregularities at the light cylinder $\rho=2$.}
\end{figure}

To arrive at a field $\psi_{\rm TSGF}$ that {\rm is} a solution, and 
that corresponds to the solution in (\ref{GF}), 
we superpose the operators ${\cal L}^{-1}_{\rm in}$
and ${\cal L}^{-1}_{\rm out}$ and take our radiation balanced solution 
to be 
\begin{equation}\label{RBsolution} 
\psi_{\rm TSGF}=\frac{1}{2}\left( {\cal L}^{-1}_{\rm in}+{\cal
L}^{-1}_{\rm out} \right)\sigma_{\rm eff}\equiv 
{\cal L}^{-1}_{\rm TSGF}\sigma_{\rm eff}.
\end{equation}
This integral equation is to be solved by iteration. From $\psi_{{\rm
TSGF},n}$, the $n^{\rm th}$ iteration for $\psi_{{\rm TSGF}}$, we
construct $\sigma_{{\rm eff},n}$, and then find the $(n+1)^{\rm th}$
iteration of the solution from
\begin{equation}\label{iteration} 
\psi_{{\rm TSGF},n+1}={\cal L}^{-1}_{\rm TSGF}\sigma_{{\rm eff},n}\ .
\end{equation}
This method is well suited to implementation as a finite difference
solution to a boundary value problem on a grid of $N$ points, like
that in Figure~\ref{coordgrid}. In this implementation, $\psi_{\rm
  TSGF}$ and $\sigma_{\rm eff}$ are one dimensional vectors of length
$N$, and ${\cal L}$, along with the chosen boundary conditions, forms
an $N\times N$ matrix. Since the form of this matrix depends on
boundary conditions, the numerical inverse ${\cal L}^{-1}$ is also
specific to the boundary conditions. The numerical radiation balanced
operator ${\cal L}_{\rm TSGF}^{-1}$ is simply the average of the
matrix inverses of ${\cal L}^{-1}$ for the ingoing and outgoing
problems. Results of this numerical solution are shown in
Figure~\ref{standwvs}.  This figure, when compared with
Figure~\ref{psiout} nicely illustrates the symmetry of the radiation
field with respect to reversal of ${\varphi}$. It is not so effective
in illustrating the fact that there is no radius at which $\psi$
vanishes at all values of ${\varphi}$. This is due to the strong
dominance of the $m=1$ multipole of the field. For larger values of
the source velocity $v=a\Omega$ the $m=3,5,\ldots{}$ multipoles make
stronger contributions, but for those values of $v$ for which we could
get accurate solutions (up to $v$ around 0.8) the $m=1$ multipole
continued to dominate the appearance of the fields in a plot like that
in Figure~\ref{standwvs}.

We now argue that within the scope of our approximation, the TSGF
solution in the wave zone is close to a linear superposition of the
ingoing and outgoing (LSIO) solutions for the same orbiting
sources. It is important to recall that due to the nonlinearity, the
LSIO is {\em not} a solution of (\ref{reducedwveq}), but it does
have the convenient property that if the nonlinearities are weak in
the wave zone, the outgoing and ingoing solutions can easily be
extracted from the LSIO\@. We are claiming then that an approximate
outgoing solution to the problem, in the wave zone, can be found from
the TSGF solution.  We emphasize that we are {\em not} claiming that
nonlinear effects are weak. We will show that this approximation
method is successful for problems with very strong nonlinear effects.
The reason for this success can be seen most easily if
(\ref{RBsolution}) is rewritten as
\begin{equation}\label{RBrewrite} 
\psi_{\rm TSGF}=\frac{1}{2}{\cal L}_{\rm in}^{-1}
\sigma_{\rm eff}(\psi_{\rm TSGF})
+ \frac{1}{2}{\cal L}^{-1}_{\rm out}\sigma_{\rm eff}(\psi_{\rm TSGF})
\end{equation}
and compared with the LSIO ({\em not} a solution of the nonlinear
problem),
\begin{equation}\label{LSIO} 
\psi_{\rm LSIO}=\frac{1}{2}{\cal L}_{\rm in}^{-1}
\sigma_{\rm eff}(\psi_{\rm in})
+ \frac{1}{2}{\cal L}^{-1}_{\rm out}\sigma_{\rm eff}(\psi_{\rm out})\ .
\end{equation}
The TSGF solution and the LSIO superposition differ due to the nonlinear
terms contained within the effective source. Those nonlinear terms
will be significant only in the small radius inner regions of the
physical space, where the fields are strong. {\em But in the inner,
strong field, regions the solution should not be highly sensitive to
whether ingoing or outgoing boundary conditions are imposed.}  Indeed,
this last statement is a way of viewing the underlying idea in our
approach.  If the fields near the orbiting objects are not
significantly influenced by the distant boundary conditions, then
radiation reaction cannot be important. We are now assuming the
converse: with parameters for which radiation reaction is not important,
the fields near the sources will not be sensitive to the boundary
conditions.

If the nonlinear contributions to $\sigma_{\rm eff}$ are nonnegligible
only in the strong field region near the orbiting objects, and if the
fields there are insensitive to boundary conditions, then we can
conclude that $\sigma_{\rm eff}(\psi_{\rm in})$ and $\sigma_{\rm
eff}(\psi_{\rm out})$ differ negligibly. It is then plausible that
they are also negligibly different from $\sigma_{\rm eff}(\psi_{\rm
TSGF})$, and hence that the LSIO is approximately the same as the TSGF
solution.  This conclusion, furthermore, should be valid to the same
degree that it is valid to ignore radiation reaction (more
specifically to ignore the sensitivity of the fields near the source
to the boundary conditions on the waves).

\begin{figure}[ht]
\begin{center}
\epsfxsize=3in\epsfbox{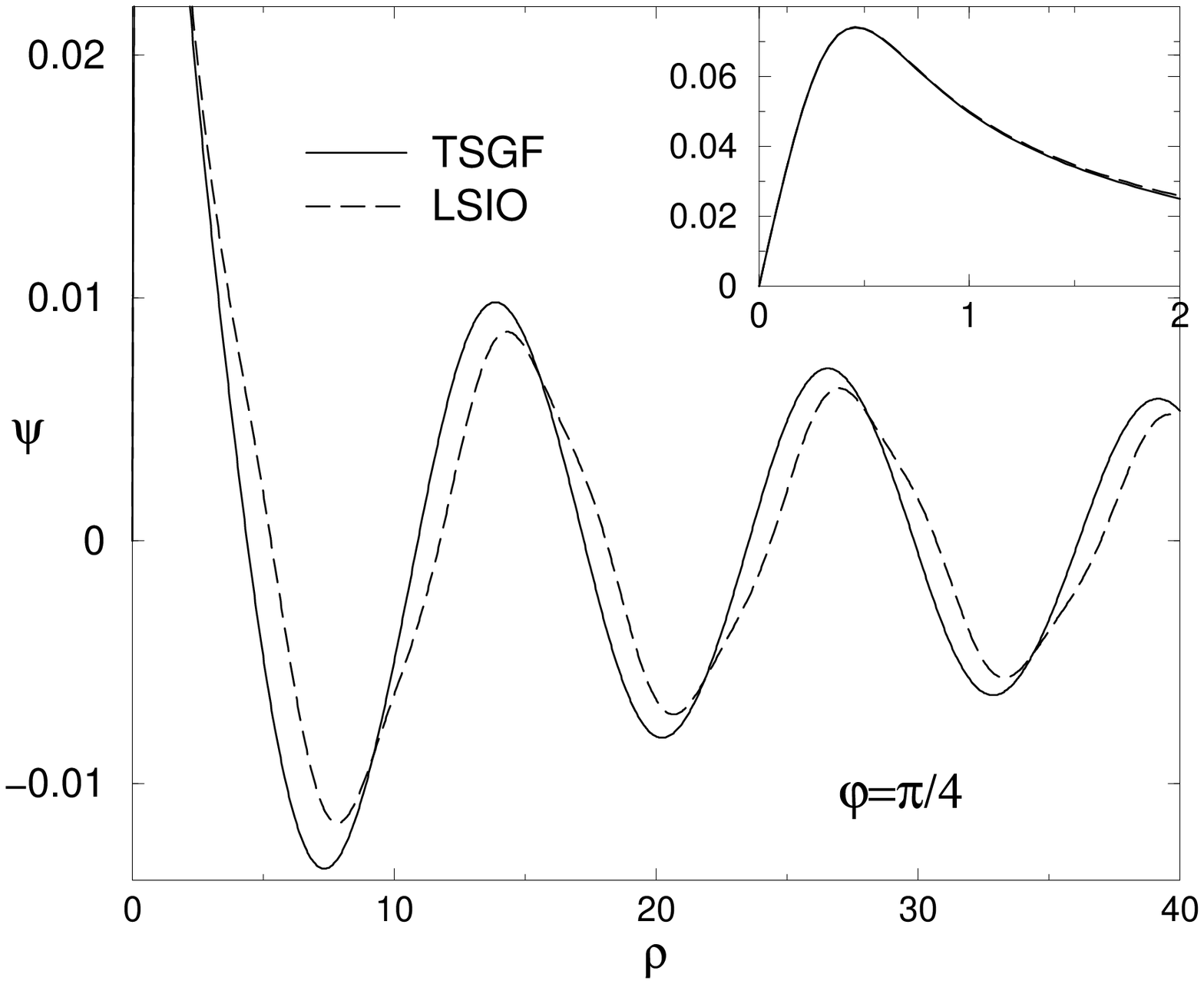}
\hspace{0.2in}
\epsfxsize=3in\epsfbox{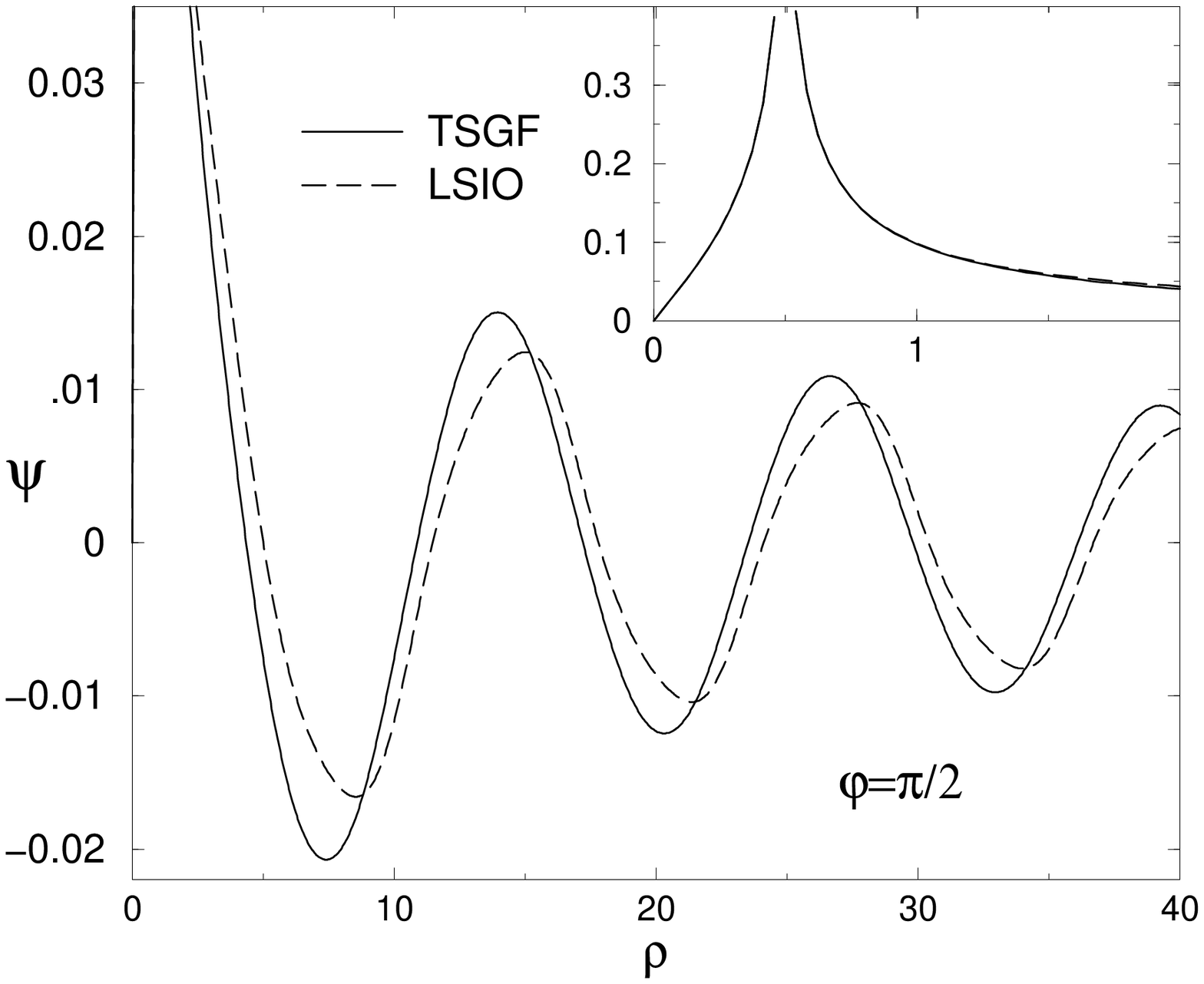}
\end{center} 
\caption[psiout]{\label{TSGFvsLSIO} The
difference between TSGF and LSIO\@.
The figures show the source region and the wave  zone
for $a=0.5$, $\Omega=0.5$, $\alpha=0.1$ and $\lambda=150$.
The plot on the right is for 
${\varphi}=\pi/4$ and that on the left is for
${\varphi}=\pi/2$, the ${\varphi}$ location of one of the particles.}
\end{figure}

For the nonlinearity given by (\ref{gaussource}), with $\lambda=1500$,
$\alpha=1$, it turns out that the approximation of TSGF by LSIO is
{\em too} good for an effective illustration. In this case a plot
shows no discernible difference between the TSGF and LSIO, even though
the nonlinearity plays a strong role, as shown in Figure~\ref{inset}.  A
discernible difference can be seen if the $\alpha$ parameter is
reduced. The effect is to spread the region of strong nonlinearity to
larger radius where the fields are somewhat sensitive to the boundary
conditions.  The comparison of the TSGF and the LSIO fields, for
$\alpha=0.1$ is shown in Figure~\ref{TSGFvsLSIO}. The field $\psi$ is
shown as a function of $\rho$ both for ${\varphi}=\pi/2$, the angular
position of a source particle, and at ${\varphi}=\pi/4$. No difference
between the TSGF and the LSIO fields can be seen at small radii. In
the radiation zone a small phase shift can be seen between the LSIO
and the TSGF and the LSIO waves are seen to have a slightly smaller
amplitude.

In our approximation approach, the idea is to find the solution of the
outgoing wave amplitude from the TSGF solution, by considering the
TSGF solution in the wave zone to be (approximately) an equal mixture
of ingoing and outgoing radiation. The specific application of this
idea requires fitting the TSGF radiation field to a sum of 
multipoles of the form
\begin{equation}
\psi_{\rm
TSGF}\approx\sum_{m=1,3,5\ldots{}}C_{m}N_{m}(m\Omega\rho)\sin{m{\varphi}},
\end{equation}
and from the amplitudes $C_{m}$ inferred from the fit, writing the
outgoing solution as
\begin{equation}
\psi_{\rm out}\approx\sum_{m=1,3,5\ldots{}}C_{m}
\left[N_{m}(m\Omega\rho)\sin{m{\varphi}}
-J_{m}(m\Omega\rho)\cos{m{\varphi}}\right]\ . 
\end{equation}
Using our toy nonlinear model, we can solve for both the TSGF and the
outgoing solution, to check the accuracy of this method.  For the
model with parameters $Q=1$, $a=0.5$, $\Omega=0.5$, $\lambda=150$,
$\alpha=0.1$, fitting the 
TSGF from 
$\rho=20$  to $\rho=40$, gives a wave amplitude for outgoing radiation
that is larger than the true amplitude by  24\%. For $\alpha=1$
the discrepancy is only 0.04\%.

\section{Conclusion and discussion}\label{sec:disc}

For our nonlinear toy scalar field model, we have demonstrated that a
radiation balanced field, the time symmetric Green function (TSGF)
solution, can be found by solving a numerical problem similar to a
boundary value problem.  Although the character of the differential
operator is elliptical inside the light cylinder and hyperbolic
outside, no special treatment of this surface was necessary, and the
solution was found with boundary value methods usually associated with
elliptical equations, despite the outer boundary being in the
``hyperbolic'' region.  We have also shown that to the extent that
radiation reaction forces can be ignored, the TSGF solution is a good
approximation to the solution with outgoing boundary conditions.

We must now ask how related methods might be brought to bear on the
problem of orbiting objects in GR. An important feature of the
nonlinear scalar toy model is that the nonlinearities in the theory do
not occur in the wave operator. This allowed us directly to recast the
problem as an integral equation by using the time-symmetric Green
function for the wave operator.  It is not at all clear how the GR
equations can be cast in a form with a linear wave operator, or
whether it is in principle impossible.  But it is not necessary that
we follow these same steps in dealing with the GR equations. What is
required, most generally, is any method for specifying a time
symmetric solution.

For such a method in GR it is expected that some features of the basic
physics will be the same as in the scalar model.  In particular, for
motions $\tau_{orb}/\tau_{rad}\ll 1$ the nonlinear terms in the
effective source should be insensitive to boundary conditions.  It can
then be supposed that the radiation balanced solution is a good
approximation to the linear superposition of ingoing and outgoing
solutions (LSIO) and that from the LSIO we can infer the outgoing
solution. There is, however, an important difference between this
claim as it applies to GR, and the (demonstrably true) claim for the
scalar field model. In the scalar field model we were comparing LSIO
fields for periodic motion with the TSGF solution for periodic motion.
In GR, there can be no LSIO solution for periodic motion, since
ingoing, or outgoing, energy flux would be incompatible with periodic
motion.  For GR, the analogy needs to be made directly to the solution
with outgoing waves and a slow rate of orbital decay due to radiation
reaction.  By constructing a spacetime with only outgoing radiation
from the TSGF solution, we arrive at an approximation in which the
radiation fields are periodic, with a period that is constant, not
slowly drifting in time.  In the notation introduced in
Section~\ref{sec:intro}, the quasi-stationary method in GR requires
that the type-III problem be solved and used to produce an
approximation to the type-I physics.

An additional complication is that the spacetime of the TSGF fields
cannot be asymptotically flat in GR, since there is an infinite amount
of energy contained in the radiation fields\cite{gibbonsstewart}. This
apparent pathology can be viewed as irrelevant if we think of our
TSGF, or outgoing, solutions as being an approximation only in a
finite central region of the space. It is worth noting that this
viewpoint is consistent with the nature of the numerical solution that
is based on boundary conditions applied at finite radius. The
important question that remains is whether the inner region, the
region in which the problem is solved, is large enough to include a
zone in which waves are weak and in which the ingoing and outgoing
waves of the approximate LSIO can be disentangled. To investigate
this, we note that the gravitational wave luminosity of the binary is
of order $L_{\rm GW} \sim (GM/ac^2)^5 (c^5/G)$, where we are using the
same notation as for (\ref{tauratio}). The gravitational wave energy
$E_{\rm GW}$ contained within a sphere of radius ${\rho_{\rm max}}$ is
of order $L_{\rm GW}{\rho_{\rm max}}/c$, and hence the ratio of
$E_{\rm GW}$ to the orbital energy $GM^{2}/a$ of the binary, is of
order
\begin{equation}
\frac{E_{\rm GW}}{GM^2/a}\sim \left( \frac{GM}{ac^{2}} 
\right)^3\frac{{\rho_{\rm max}}}{a}\ .
\end{equation}
The nature of our approximation requires that $GM/(ac^{2})$ be small,
so we conclude that the gravitational wave energy contained within
${\rho_{\rm max}}$ will be much less than the orbital energy, even for
values of ${\rho_{\rm max}}$ large compared to $a$.

This last conclusion is important if we are to hope to use versions of
our method to make inferences about the ISCO\@. By investigating the
dependence of the binary energy on radius, we can study whether there
is an instability like that for particles. The ``energy'' of the orbit
must be computed as a surface integral at a large radius. If this
integral were significantly influenced by the energy contained in the
waves, conclusions about orbital stability would be suspect.

Whether or not the methods of this paper can be applied to the ISCO
question, the use of these methods in GR would provide
an important addition to the tools needed to understand black hole
inspiral.  A direct and obvious use of the method would be to provide
Cauchy data for numerical relativity. 

\ack


We would like to thank Steven Detweiler for very helpful discussions.
We thank also Patrick Brady, Teviet Creighton, \'{E}anna Flanagan,
Scott Hughes, Kip Thorne, and Alan Wiseman for useful discussions at
meetings on the intermediate black hole problem at Caltech.  We are
grateful also to Ji\v{r}\'{\i} Bi\v{c}\'{a}k and John Friedman for
helpful suggestions.  This work was partially supported by the
National Science Foundation under grant PHY9734871. One of us (JTW)
acknowledges support by the Swiss Nationalfonds, and by the Tomalla
Foundation, Z\"{u}rich.

\appendix

\section{General Radiation-Balanced Solution}\label{app:gfsoln}

We present here some of the details of the general solution for the
linear scalar field theory with equal flux of ingoing and outgoing
radiation. The Green function $G(\rho,{\varphi};\rho',{\varphi}')$ for
the linear problem satisfies
  \begin{equation}
    \fl\left[\frac{1}{\rho}\frac{\partial}{\partial\rho}
      \rho\frac{\partial}{\partial\rho}
      +\left(
        \frac{1}{\rho^{2}}-\Omega^{2}
      \right)
      \frac{\partial^{2}}{\partial{\varphi}^{2}}
    \right]G(\rho,{\varphi};\rho',{\varphi}')
    =-\rho'^{-1}\delta(\rho-\rho')\delta({\varphi}-{\varphi}').
  \end{equation}
When the ${\varphi}$ dependence is represented in a Fourier series, and
the Green function is written as $G(\rho,{\varphi};\rho',{\varphi}')$
=Re$\left[\sum{\cal
G}_m(\rho;\rho',{\varphi}')\rme^{im{\varphi}}\right]$\,,
the equation becomes 
  \begin{equation}\label{generalRB} 
\left[\frac{1}{\rho}\frac{\partial}{\partial\rho}
\rho\frac{\partial}{\partial\rho}+m^2
\left(
\Omega^{2}-
\frac{1}{\rho^{2}}
\right)
\right]
{\cal
G}_m(\rho;\rho',{\varphi}')
=-\frac{1}{2\pi\rho'}
\delta(\rho-\rho')\rme^{-im{\varphi}'}
  \end{equation}
Reality of the Green function $G(\rho,{\varphi};\rho',{\varphi}')$
means that
${\cal G}_{-m}(\rho;\rho',{\varphi}')={\cal
  G}_m(\rho;\rho',{\varphi}')^*$, and the $m=0$ mode is irrelevant
for the sources we consider, so we only need the general solution to
(\ref{generalRB}) for $m>0$, which is
\begin{equation}
\fl
{\cal G}_m(\rho;\rho',{\varphi}')=\cases{
  -\frac{1}{4}
    J_m(m\Omega \rho')N_m(m\Omega\rho)\rme^{-im{\varphi}'}
+\Gamma_m(\rho',{\varphi}')J_m(m\Omega\rho)
     & $\rho>\rho'$\\
%
%
     -\frac{1}{4}
    N_m(m\Omega \rho')J_m(m\Omega\rho)\rme^{-im{\varphi}'} 
+\Gamma_m(\rho',{\varphi}')J_m(m\Omega\rho)
    & $\rho<\rho'$
}
\end{equation}
For $\rho>\rho'$, ${\cal G}_m$
can be written in terms of Hankel functions as
\begin{equation}
{\cal G}_m=A_mH^{(1)}_m(m\Omega\rho)+B_mH^{(2)}_m(m\Omega\rho)\ ,
\end{equation}
where 
\begin{eqnarray}
A_m&=&\frac{1}{2}\Gamma_m(\rho',{\varphi}')
-\frac{1}{8i}J_m(m\Omega\rho)\rme^{-im{\varphi}'}
\nonumber\\
B_m&=&\frac{1}{2}\Gamma_m(\rho',{\varphi}')
+\frac{1}{8i}J_m(m\Omega\rho)\rme^{-im{\varphi}'}\label{AandB}  \ .
\end{eqnarray}
The condition for equal flux of ingoing and outgoing radiation is
$|A_m|=|B_m|$, for all $m$. This, and the expressions in
(\ref{AandB}), require that $\Gamma_m(\rho',{\varphi}')$ be a real
multiple of $J_m(m\Omega\rho)\rme^{-im{\varphi}'}$.  The
proportionality constant may be different for each $m$, so the general
radiation balanced solution is
\begin{eqnarray}
\fl
{\cal G}_m(\rho;\rho',{\varphi}')=
\cases{
  -\frac{1}{4}
    J_m(m\Omega \rho')N_m(m\Omega\rho)\rme^{-im{\varphi}'}
+g_mJ_m(m\Omega \rho')J_m(m\Omega\rho)\rme^{-im{\varphi}'}
     & $\rho>\rho'$\\
%
%
     -\frac{1}{4}
    N_m(m\Omega \rho')J_m(m\Omega\rho)\rme^{-im{\varphi}'}
+g_mJ_m(m\Omega \rho')J_m(m\Omega\rho)\rme^{-im{\varphi}'}
    & $\rho<\rho'$
}
\nonumber\\
\end{eqnarray}
Any choice of the set of real constants $g_m$ gives a radiation
balanced solution. The choice made for the TSGF solution, $g_m=0$, for
all $m$, is convenient for a numerical implementation that does not
explicitly use Fourier decomposition.  The choice $g_m=0$ also seems
to be the most natural one if one views the $\Gamma_m$ terms in
(\ref{generalRB}) as free waves, disconnected from the source.

\section{Three-plus-One Dimensional Theory}\label{app:threeD}
Here we present the solution of a linear toy
model for point-like sources.  We use standard spherical coordinates
($r,\theta,\phi$) and suppose that two particles of opposite scalar
charge $\pm Q$ orbit at frequency $\Omega$ in the equatorial plane
($\theta=\pi/2$), at radius $a$, with angular separation
$\Delta\phi=\pi$. In the linear differential equation of this toy
model
\begin{equation}\label{genlin}
\fl\nabla^{2}\psi-\partial_{t}^{2}\psi=
-a^{-2}Q\delta(r-a)\delta(\cos\theta)
\left[\delta(\phi-\Omega t-\pi/2)-\delta(\phi-\Omega t-3\pi/2) \right]
\ ,
\end{equation}
we make the {\it ansatz} that the solution is not of the general type
$\psi(r,\theta,\phi,t)$, but rather of the type
$\psi(r,\theta,\varphi)$ where $\varphi\equiv\phi-\Omega t$. The
differential equation then reduces to the point-like linear analog of
(\ref{reducedwveq}):
\begin{eqnarray}
\fl\frac{1}{r^2}\frac{\partial}{\partial r}\left(
r^2\frac{\partial\psi}{\partial r} \right)
+\frac{1}{r^2\sin\theta}\frac{\partial}{\partial\theta}\left(
\sin\theta\frac{\partial\psi}{\partial\theta}
\right)
+\left[
\frac{1}{r^{2}\sin^2\theta}-\Omega^{2}
\right]\frac{\partial^{2}\psi}{\partial{\varphi}^{2}}
\nonumber\\
=a^{-2}Q\delta(\rho-a)\delta(\cos\theta)
\left[\delta({\varphi}-\pi/2)-\delta({\varphi}-3\pi/2) \right]\ .
\label{ptlike}
\end{eqnarray}
If $\psi(r,\theta,\varphi)$ is decomposed as a sum of spherical
harmonics $Y_{\ell m}(\theta,\varphi)$ it is straightforward to find
the solution that is well behaved at $r=0$ and that corresponds to
outgoing waves at infinity. This solution, the point-like equivalent
of (\ref{Bessum}), is
\begin{equation}\label{ptsoln}
\psi(r,\theta,\varphi)
=Q\Omega\sum_{\ell m}\kappa_{\ell m}Y_{\ell m}(\theta,\varphi)   
\times\cases{
    j_\ell(m\Omega a) h^{(1)}_\ell(m\Omega r) & $r>a$\\
%
%
    j_\ell(m\Omega r) h^{(1)}_\ell(m\Omega a)      &$r<a$
}
\end{equation}
Here $j_\ell$ and $h^{(1)}_\ell$ indicate spherical Bessel and Hankel
functions. The sum in (\ref{ptsoln}) is over odd $\ell$ and odd $m$,
and the coefficients $\kappa_{\ell m}$ are given by
      \begin{equation}
\fl\kappa_{\ell m}= 
im\left[Y_{\ell m}^{*}(\pi/2,\pi/2)-Y_{\ell m}^{*}(\pi/2,3\pi/2)\right]
=(-1)^{(m-1)/2}2mY_{\ell m}(\pi/2,0)
\ .
      \end{equation}
The point-like equivalent of (\ref{halfoutin}), the radiation balanced
TSGF solution of the problem, is
\begin{equation}\label{ptsolnmerb}
\psi(r,\theta,\varphi)
=iQ\Omega\sum_{\ell m}\kappa_{\ell m}Y_{\ell m}(\theta,\varphi)
\times\cases{
    j_\ell(m\Omega a) n_\ell(m\Omega r) & $r>a$\\
%
%
    j_\ell(m\Omega r) n_\ell(m\Omega a)      &$r<a$
}
\end{equation}
where $n_\ell$ is the spherical Neumann function.  These solutions
serve as explicit illustrations that for point-like sources in 3+1
dimensions the field embodies the same symmetry as the sources. That
is, the field rotates ``rigidly.'' The dependence on time and on
azimuthal angle appears only in the combination $\phi-\Omega t$.

\section{Net Force on the Orbiting Particles}\label{app:force}

In this appendix, we show that for a general radiation-balanced
solution to the problem of orbiting particles, the proscribed circular
orbits of the particles are consistent with the scalar field equations
of motion without the need for external forces.  (This is not true in
the case of purely outgoing radiation.)  We limit attention to the
linear theory, so that we can easily separate out the forces due to
each particle on the other, without worrying about any self-force.

The covariant force law for a scalar-charged particle with charge $Q$,
mass $m$, and energy-momentum vector $p$ moving in a scalar field
$\psi$ gives a force of
\begin{equation}
 m^{-1} p^\nu \nabla_\nu p^\mu = \frac{\rmd p^\mu}{\rmd\tau} 
+ m^{-1} \Gamma^\mu_{\nu\lambda} p^\nu p^\lambda
= -Q \nabla^\mu \psi
\ .
\end{equation}
If we work in co-rotating coordinates, in which the particle is
momentarily at rest, the metric is
\begin{equation}
  \rmd s^2 = -(1-\Omega^2\rho^2)\rmd t^2
  +\rmd\rho^2+\rho^2\rmd\varphi^2
  +2\Omega\rho^2 \rmd\varphi\rmd t
\end{equation}
and the only non-zero component of the energy-momentum vector is
\begin{equation}
  p^t = (1-\Omega^2a^2)^{-1/2} m =: \gamma m
\ ,
\end{equation}
so that the only relevant Christoffel symbol is $\Gamma^\rho_{tt}$.
(In the corresponding calculation in the 3+1 dimensional theory, only
$\Gamma^r_{tt}$ is relevant.)

The equations of motion for $p^\rho$ and $p^\varphi$ in the 2+1
dimensional theory thus become
\begin{eqnarray}
  \label{radforce}
  \frac{\rmd p^\rho}{\rmd\tau} 
  = - Q \psi_{,\rho} + \gamma^2m\rho\Omega^2\\
  \label{angforce}
  \frac{\rmd p^\varphi}{\rmd\tau} 
  = - Q g^{\varphi\varphi} \psi_{,\varphi}
\ ,
\end{eqnarray}
where we have used the form of the inverse metric $g_{\mu\nu}$ in
co-rotating coordinates and the fact that $(\partial\psi/\partial
t)_\varphi=0$.  The second term in (\ref{radforce}) represents the
fictitious centrifugal force that the particle feels in the
co-rotating reference frame.  The radial momentum $p^\rho$ can remain
zero if the repulsive effect of this term cancels the attraction due
to the field, described in the first term.  This implies a
relationship among the orbital radius $a$, angular frequency $\Omega$,
and charge-to-mass ratio $Q/m$.  (The corresponding relationship is
given for orbiting charged particles in the presence of a
half-advanced, half-retarded electromagnetic potential in
\cite{schild}.)

The angular force in (\ref{angforce}), which represents radiation
reaction force, will vanish if (and only if) the scalar field due to
one particle has vanishing $\varphi$ derivative at the location of the
other particle.  For concreteness, we consider the force on the
positively-charged particle at $\rho=a$, $\varphi=\pi/2$ due to the
negatively-charged particle at $\rho=a$, $\varphi=3\pi/2$.  The field
due to the latter particle is
\begin{equation}
  \psi(\rho,\varphi) = -Q G(\rho,\varphi;a,3\pi/2)
  \ ,
\end{equation}
where $G(\rho,\varphi;\rho',\varphi')$ is the Green function used to
construct the solution.  Using the Fourier expansion of the Green
function given in \ref{app:gfsoln}, we find that
\begin{equation}
  \fl
  \psi_{,\varphi}(a,\pi/2) = -{\rm Re}
  \left[
    \sum_m im
    \left(
      \frac{1}{4}J_m(m\Omega a)N_m(m\Omega a)(-1)^m
      +\Gamma_m(a,3\pi/2)\rme^{im\pi/2}
    \right)
  \right]
  \ .
\end{equation}
This vanishes if the quantity in large parentheses on the right-hand
side is real; for a general radiation-balanced solution this is the
case, since $\Gamma_m(a,3\pi/2)=g_m J_m(m\Omega a) \exp(-im3\pi/2)$
with $g_m$ real.  Thus \emph{the radiation-balanced solutions have
  zero radiation reaction force}.

However, in the case of purely outgoing radiation,
$\Gamma_m(\rho',\varphi')=J_m \rme^{-im\varphi'}/4i$, and we can
explicitly find the radiation reaction force as
\begin{equation}
  \psi_{,\varphi}(a,\pi/2) 
  = -\sum_m (-1)^m\frac{m}{4} [J_m(m\Omega a)]^2 \ne 0
  \ .
\end{equation}

A similar demonstration can be made in the 3+1-dimensional case using
the outgoing-radiation and radiation-balanced fields (\ref{ptsoln})
and (\ref{ptsolnmerb}), using the symmetries of the spherical Bessel
functions under changes of sign in their arguments to illustrate that
terms in the radiation-reaction force due to terms with opposite signs
of $m$ cancel each other out in the radiation-balanced mode expansion
but not in the outgoing-radiation one.

\end{document}